\documentclass[11pt]{article}
\usepackage[english]{babel}

\usepackage[a4paper,top=2cm,bottom=2cm,left=3cm,right=3cm,marginparwidth=1.75cm]{geometry}

\usepackage{scalerel,stackengine,amsmath}
\usepackage{graphicx, setspace}
\usepackage{caption}
\usepackage{subcaption}
\usepackage{comment}
\usepackage{authblk}
\usepackage{gensymb}
\usepackage{multirow}

\usepackage{soul}
\usepackage{csquotes}
\usepackage{dirtytalk}

\usepackage{booktabs}
\usepackage[colorlinks=true, allcolors=blue]{hyperref}
\usepackage[style=chem-angew,sorting=none, autocite = superscript]{biblatex}

\let\cite\autocite

\providecommand{\keywords}[1]
{
  \small	
  \textbf{\textit{Keywords---}} #1
}

\title{Crystallographic Control of Hydrogen Ingress in Bcc-Iron: Insights from ab initio Simulations}
\author[1,2]{Lukas Meier\thanks{lukas.meier@ugent.be}}
\author[1,2]{Asif I. Bhatti}
\author[2]{Leo Kestens}
\author[1,2]{Stefaan Cottenier\thanks{stefaan.cottenier@ugent.be}}

\affil[1]{Center for Molecular Modeling, Ghent University, Technologiepark 46, BE-9052 Zwijnaarde, Belgium}
\affil[2]{Department of Electromechanical, Systems and Metal Engineering, Ghent University}

\begin{document}

\maketitle

\begin{abstract}
Hydrogen uptake into body-centered cubic (bcc) iron as a root cause for subsequent hydrogen embrittlement, is initiated at the surface. In this paper, we quantify how readily H diffuses from the surface into the bulk. We consider a set of low-index, vicinal and general Fe surfaces and treat H-permeation as a two-step process. First, density-functional calculations determine the adsorption energy of an isolated H atom at every crystallographically distinct surface site. Second, for each adsorption site we map the minimum-energy pathway that carries the atom beneath the surface and into the lattice. Across all ten orientations studied, a clear trend emerges: sites that bind hydrogen most weakly (highest adsorption energy) are the starting point of the lowest-barrier diffusion channels into the metal interior. Thus, the least-favorable adsorption pockets act as “gateways” for efficient subsurface penetration. These insights provide a practical design rule: suppressing or minimizing exposure of such high-energy adsorption motifs - through appropriate surface texturing or orientation control - should make bcc-iron components less susceptible to hydrogen uptake and the associated embrittlement.
\end{abstract}


\section{Introduction}\label{sec:intro}
With the accelerated effects of the climate crisis, hydrogen is expected to play an increasingly important role as an energy carrier. This poses new challenges on the existing infrastructure and places the interaction between hydrogen and steel at a critical frontier.
Specifically, the hydrogen-induced deterioration of metals, known as hydrogen embrittlement (HE), is one of the biggest roadblocks in this regard \cite{Djukic2019, Matsumoto2009, Li2020_Review_HE}.
To mitigate the effects of HE, there are two major ways: i) increase the material's internal resistance against HE by designing improved materials with a different microstructure or alloying composition, or ii) reduce the amount of hydrogen that can enter the material through its surface, which is the road that this paper addresses.
Under normal conditions, most steel surfaces are not directly exposed to hydrogen, as they either possess a natural oxide layer or - depending on the application - some kind of coating that partially protects them against hydrogen entry. However, under working conditions and high pressures, microcracks can form and expose the metallic surface directly to the hydrogen gas.
In this context, it is relevant to investigate, how the metallic steel surface should ideally look like in order to minimize hydrogen entry \cite{Zhu2024, Laadel2022, Li2022}.

In this regard, there are several parameters to optimize: The steel's alloy composition, its microstructure, as well as its texture.
In this study, we focus on the effect of the texture, i.e., the orientation of the monocrystalline grains at the surface. Therein, we study pure bcc-iron as the precursor material to steel and investigate, how and where atomic hydrogen adsorbs on bcc-surfaces with different orientations. With the surface adsorption sites as starting points for the diffusion process, we investigate in a second step, how easy it is for hydrogen to diffuse into the iron bulk region.

Specifically, we further test the hypothesis raised in our recent review \cite{Meier2024}: Do general surface sites inherit their adsorption characteristics from one of the low-index surfaces (100), (110), and (111)? And if that is the case, can we predict hydrogen adsorption characteristics on a given surface site purely based on the surface site's geometric similarity with one of the low-index surfaces?

In a second part, we calculate the diffusion profile from the different investigated surfaces up to the first stable subsurface sites. Together with our results on hydrogen adsorption, we formulate a geometric criterion to determine, which surfaces possess \say{easy} entry points into the material and should therefore be minimized in a hydrogen-resilient texture.
\\

Although the problem of hydrogen entering steel is widely known and the hydrogen/iron adsorption and diffusion has been studied using \emph{ab initio} methods for two decades, previous studies are mostly limited to the low-index surfaces, among which especially the (100) and (110) surfaces have been extensively studied \cite{Meier2024, Jiang2004, Sorescu2005, Boda2019, Shen2016, Li2020, Li2019_2}.
This leaves the question open to how general high-index surfaces behave. 

\begin{figure}
    \centering
    \includegraphics[width=0.5\linewidth]{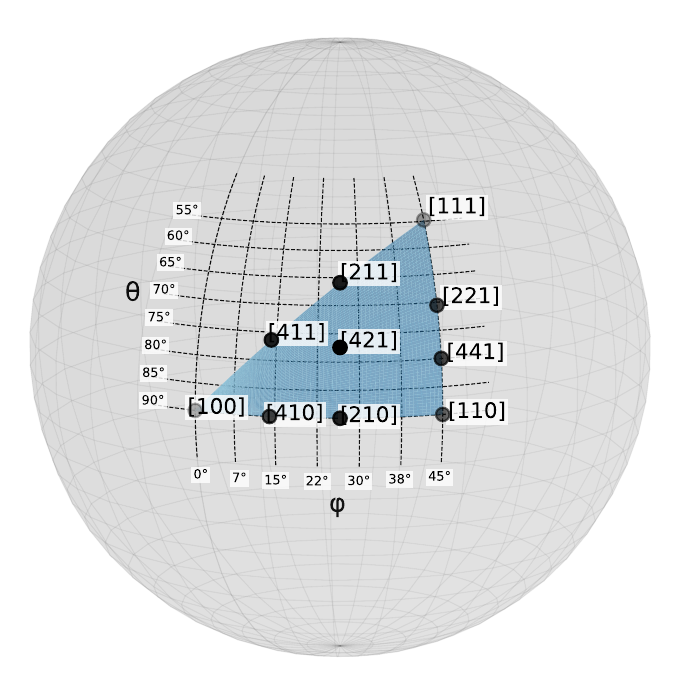}
    \caption{Geometries of the investigated surfaces represented by the intersection of their surface normal vectors with the unit sphere. The polar and azimuth angles as defined in Tab.~\ref{tab:surface_geometries} are indicated.}
    \label{fig:triangle_black}
\end{figure}

In our study we included - besides the low-index surfaces - 7 different other surfaces, whose polar coordinates are given in Tab.~\ref{tab:surface_geometries} and which are displayed in Fig.~\ref{fig:triangle_black} (For further information regarding the surface geometries and the relation between the angles given in Tab.~\ref{tab:surface_geometries} and the Miller Index representation, see \ref{appx:geometry}).

\begin{table}
    \centering
    \begin{tabular}{c|c|c|c|c|c|c|c|c|c|c}
        Surface & (100) & (410) & (210) & (110) & (411) & (421) & (441) & (211) & (221) & (111)\\
        \hline
        $\varphi [\degree]$ & 0 & 14 & 27 & 45 & 14 & 27 & 45 & 27 & 45 & 45\\
        $\vartheta [\degree]$ & 90 & 90 & 90 & 90 & 76 & 77 & 80 & 66 & 71 & 55\\
    \end{tabular}
    \caption{All surfaces investigated with their corresponding geometrical properties expressed by the azimuth angle ($\varphi$) and the polar angle ($\vartheta$) enclosed by their surface normal with the [100] direction. The computational details to model these surfaces can be found in \ref{appx:comp_details}}
    \label{tab:surface_geometries}
\end{table}

The present set of surfaces samples the whole space of surface orientations given by the pair of polar and azimuth angles in chunks of about 10 - 15$\degree$, ensuring an equal sampling spread. An even finer sampling would require surfaces with larger miller indices and therefore larger unit cells, leading to considerably longer simulation times.

Another distinct feature of some surfaces is that they form terraces resembling one of the low-index surfaces with steps in between. These surfaces are so-called \emph{vicinal} surfaces and the spatial extension of their terrace depends on the angle enclosed between the surface normal and the terrace's surface normal. In the limit of the enclosing angle approaching zero, the terraces become infinitely large, resulting in the respective low-index surface.
This being said, the required \say{minimum} angle to resemble locally a low-index plane geometry differs depending on the lateral dimensions of the respective low-index surface's unit cells. With the (100) and (110) surfaces having a laterally smaller unit cell than the (111) surface with its more open structure, it is even possible to identify (100) and (110) vicinal planes on the (210) and (221) surfaces. On the other hand, it is barely possible to identify a (111) vicinal plane on the (221) surface, as the (111) unit cell requires a significantly larger area, see \ref{appx:comp_details}, Tab.~\ref{tab:unitcell_measures}

\section{Methodology}
Our study is based on density functional theory (DFT) as implemented in the Vienna ab initio simulation package (VASP) \cite{Kresse1996}. The electron-ion interaction is described by the projector-augmented wave (PAW) scheme \cite{Kresse1999}. The generalized gradient approximation (GGA) in form of so-called GW-ready potentials is used as approximation to the exchange-correlation functional. These have shown superior agreement with high-precision all-electron calculations \cite{Bosoni2023}. Specifically, we have chosen the \emph{PAW H\_GW 21Apr2008} and \emph{PAW Fe\_GW 31Mar2010} potentials for hydrogen and iron, respectively.
To account for the effect of magnetism in bcc-iron, we incorporated the spin polarization within the colinear approximation. All spin-polarized calculations are initialized in the ferromagnetic state.
The electronic wavefunctions are expanded up to a kinetic energy cutoff of 400 eV.

\subsection{Surface slab modeling and Brillouin zone integration}
All surfaces are modeled as cuboid unit cells and were generated using the Atomsk and ASE codes \cite{Hirel2015,Larsen2017}.

This study aims at investigating \emph{single} hydrogen adsorption and diffusion through different surfaces. Thus, the surface unit cells should be sufficiently large to exclude a self-interaction between periodic copies of hydrogen. To strike a balance between realism and computational efficiency, the lateral distance between the adsorbing hydrogen and its periodic image is at least $\sqrt{2}a_0 = 4$~\AA. At this distance, no relevant interaction between two hydrogen atoms in the bulk is found, anymore \cite{Hayward2013}.

\begin{figure}
    \centering
    \includegraphics[width=0.5\linewidth]{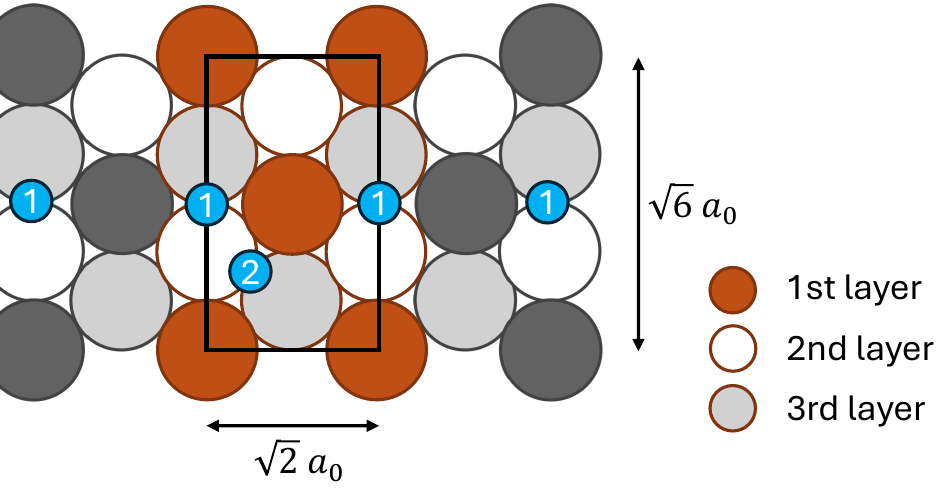}
    \caption{Schematic top-view of the studied (111) unit cell. First, second, and third layer iron atoms are displayed in brown, white, and gray, respectively. Within this unit cell there are two options to place a hydrogen atom at the deep bridge (DB) site. At site (1) the adsorption energy is artificially increased by 0.07~eV compared to site (2) as hydrogen binds to the central surface layer iron atom twice via its periodic copy. }
    \label{fig:periodic_copy_effect}
\end{figure}

This being said, a word of caution is required when describing adsorption geometries at which hydrogen would bind to the same iron atom twice (once in the unit cell and once to its periodic copy) forming an infinite linear chain. This will artificially alter the adsorption energy compared to the dilute case, which we aim to describe here, see Fig.~\ref{fig:periodic_copy_effect}. Specifically, bridge site configurations on the surface, where two adjacent iron atoms are no more than $\sqrt{2} a_0$ apart from each other are prone to this effect, which requires special diligence. When calculating the adsorption energies of hydrogen at local minima at surfaces where this becomes relevant (in our case, this was for the (111) and (221) surface), we use an equivalent site on the surface without that problem (see Fig.~\ref{fig:periodic_copy_effect}) or repeat our calculations using double the size of the unit cell in the respective direction while keeping all other computational parameters unchanged. For the calculations of the potential energy surfaces as a first approximation of adsorption minima locations, this correction was not taken into account.

\begin{figure}[tbh] 
  \begin{subfigure}[b]{0.49\textwidth}
    \includegraphics[width=\textwidth]{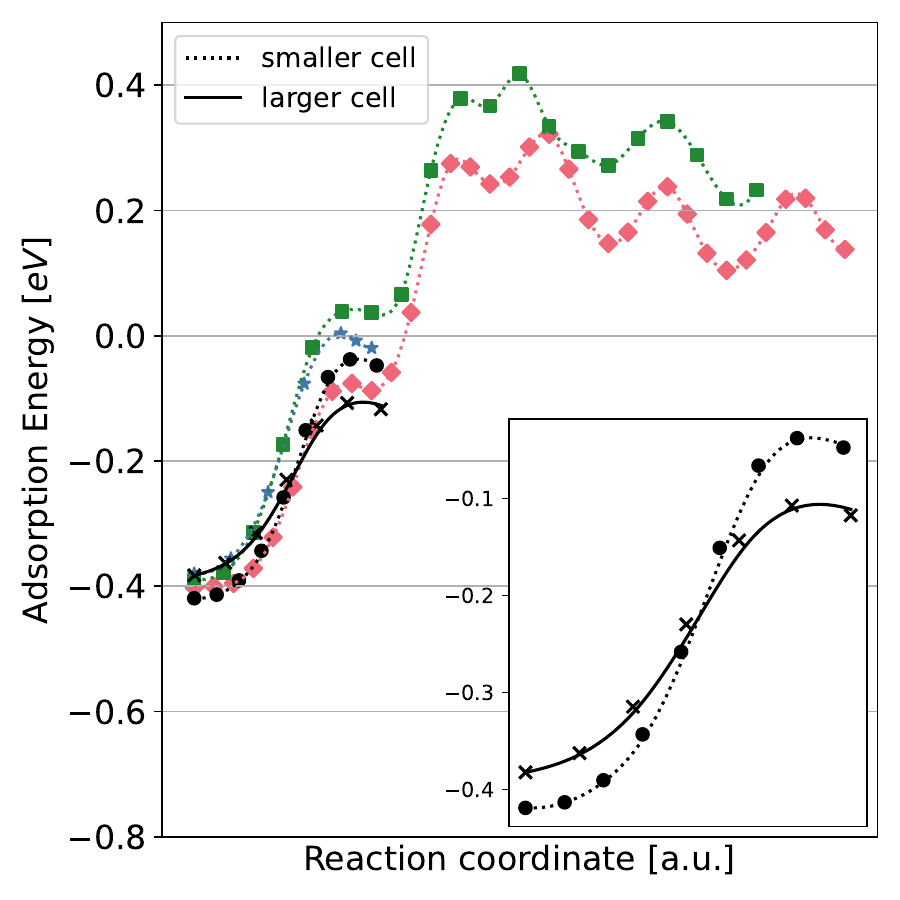}
    \caption{(100)}
    \label{fig:100}
  \end{subfigure}
  \hfill
  \begin{subfigure}[b]{0.49\textwidth}
    \includegraphics[width=\textwidth]{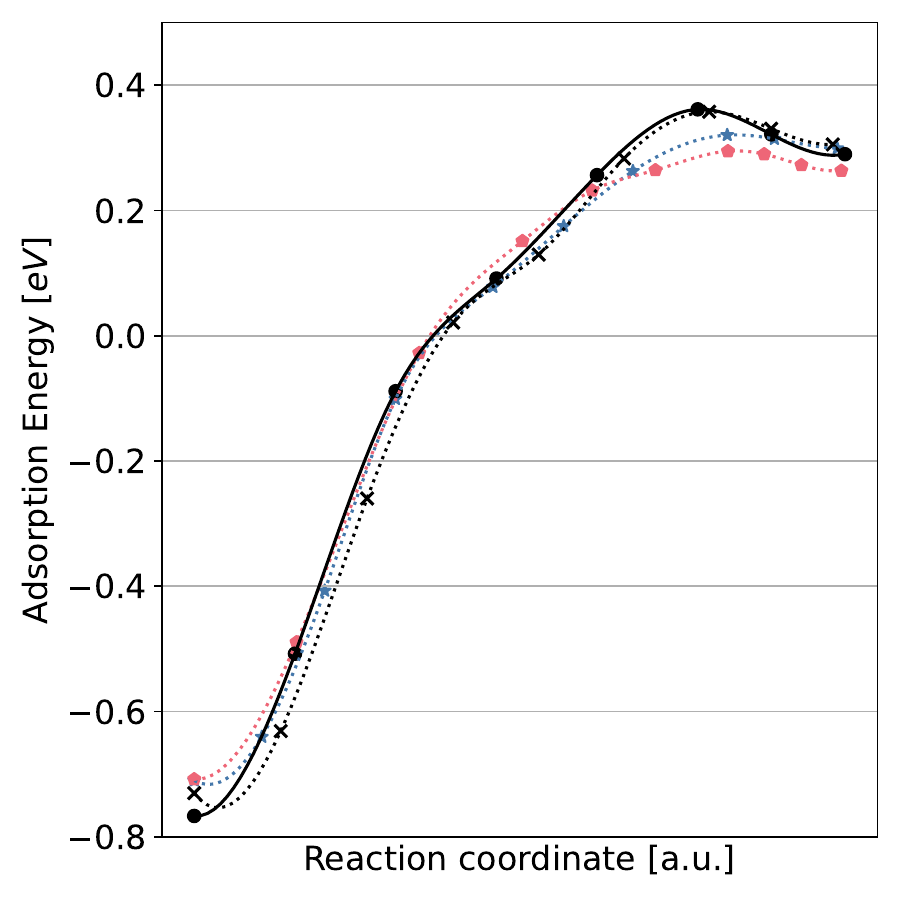}
    \caption{(110)}
    \label{fig:110}
  \end{subfigure}
  \caption{Comparison between our calculated diffusion profiles (black) and diffusion profiles from literature: \textbf{a) (100)} \cite{Jiang2004} (blue, star), \cite{Boda2019} (green, squares), and \cite{Sorescu2005} (red, diamonds). \textbf{b) (110)} \cite{Li2020} (red, pentagons). The black dashed diffusion profiles (circles) correspond to our calculations for smaller unit cells as given in Tab.~\ref{tab:surface_geometries} The solid lines (cross symbols) correspond to single hydrogen diffusion for laterally larger cell sizes of ($3a_0$, $3a_0$) and ($2\sqrt{2}a_0$, $2a_0$ ) for (100) and (110), respectively.}
  \label{fig:100-110_comp}
\end{figure}

To obtain a quantitative estimate of the error induced by the lateral size of the unit cell, we calculated the diffusion profiles for hydrogen through the (100) and (110) unit cells for different unit cell sizes. Despite the excellent qualitative agreement with earlier works, the restricted unit cell sizes used in our simulations lead to error bars of up to 0.04 and 0.1~eV for adsorption energy and diffusion barrier, see Fig.~\ref{fig:100-110_comp}
Nevertheless, we do not expect the general conclusion of this work to be affected. This is because the largest deviation stems from a lowering of the (100) diffusion barrier for a larger cell, while on the other hand, the diffusion profile through the (110) surface is even slightly increased when increasing the unit cell size. This only emphasizes our later results pointing towards (100)-like surfaces as \say{easy} pathways for hydrogen surface-to-subsurface diffusion.

Next to the lateral size, the relative adsorption energies for different surfaces depend on the thickness of the surface slabs and the number of iron layers that can freely relax. To account for the bulk-like behavior, we modeled each surface slab with a minimum thickness of $4a_0$, where $a_0$ is the bcc iron lattice constant. The lower layers which contribute to a thickness of around $1a_0$ are frozen at their respective bulk positions. Due to the different interlayer distances for different surface orientations, the total number of iron layers can vary. In the case of the (100) surface, this corresponds to a total of nine layers, with the bottom three layers frozen. 
A vacuum layer of 1~nm is used to decouple the layers along the surface normal and create a vacuum interface.
The dimensions of each unit cell, as well as the number of frozen layers are given in Sec.~\ref{appx:comp_details}, Tab.~\ref{tab:unitcell_measures}. 

Brillouin zone integration was performed with $\Gamma$-centered \textit{k}-point meshes with a density of $\left(\frac{2}{a}~\times~\frac{2}{b}~\times 1 \right)$, rounded up where necessary. $a$ and $b$ refer to the lateral dimensions of the simulated unit cells in units of the lattice constant $a_0$ (i.e., the first two columns of Tab.~\ref{tab:unitcell_measures}). For the (100), (110), and (111) surfaces serving as reference systems, the selected \textit{k}-point mesh and plane-wave cutoff result in an error bar of less than 20~meV for adsorption energies and forces of less than 20~meV/Å acting on each atom compared to higher settings. For the electronic density of states (DOS) calculations, we increased the \textit{k}-point mesh density to $\left(\frac{15}{2a}~\times~\frac{15}{2b}~\times 1 \right)$. The first-order Methfessel-Paxton method \cite{Methfessel1989} with a smearing parameter of 0.1 eV was used for the Brillouin zone integration.

\subsection{PES and NEB calculation}

To get a comprehensive overview of the possible adsorption sites and the hydrogen adsorption profile, in a first step we calculated the Potential Energy Surface (PES) of hydrogen on top of the various iron surface slabs.
Depending on the specific iron surface, the lateral separation distance of the PES sampling points differs to specifically sample high-symmetry points on the surfaces. At these sampling points, the hydrogen coordinates lateral to the surface plane are frozen, while all other degrees of freedom are allowed to fully relax. 
For the vicinal surfaces, we further calculated the PESs parallel to the low index planes to allow for a better comparison with the respective low index surfaces. A detailed summary of PES sampling grids can be found in Tab. \ref{tab:PES_sampling}. 
 
To reduce the computational effort for obtaining the PESs, each structure was pre-relaxed using the molecular dynamics simulator LAMMPS \cite{Plimpton1995} with an embedded atom potential \cite{Wen2021}. Afterwards, the structures were relaxed with VASP using a force threshold for the relaxations of 0.05~eV/\AA. For all successive quantitative relaxation calculations (adsorption energies, NEB), the force-threshold for ionic relaxations was tightened to 0.02~eV/\AA.

The adsorption energies on the surface and dissolution energies on the material are calculated via

\begin{eqnarray} \label{eq:adsorption_energy}
    E_{ads} = E_{tot}-E_s-\frac{1}{2}E_{H_2}.
\end{eqnarray}

Here, $E_{tot}$ as the energy of the combined hydrogen-iron adsystem, $E_s$ the energy of the pure iron surface slab, and $E_{H_2}$ the DFT-calculated energy of an isolated H$_2$ molecule.

Starting from the PES-obtained adsorption minima, we performed nudged elastic band (NEB) calculations \cite{Henkelman2000} to obtain energy barriers for hydrogen diffusion into the material. More precisely, a force-based conjugate gradient optimizer together with the climbing image algorithm was used. The workflow is exemplarily shown in Fig.~\ref{fig:ads_pes_scheme}

\section{Adsorption} \label{sec:adsorption}

\begin{figure}
    \centering
    \includegraphics[width=1\linewidth]{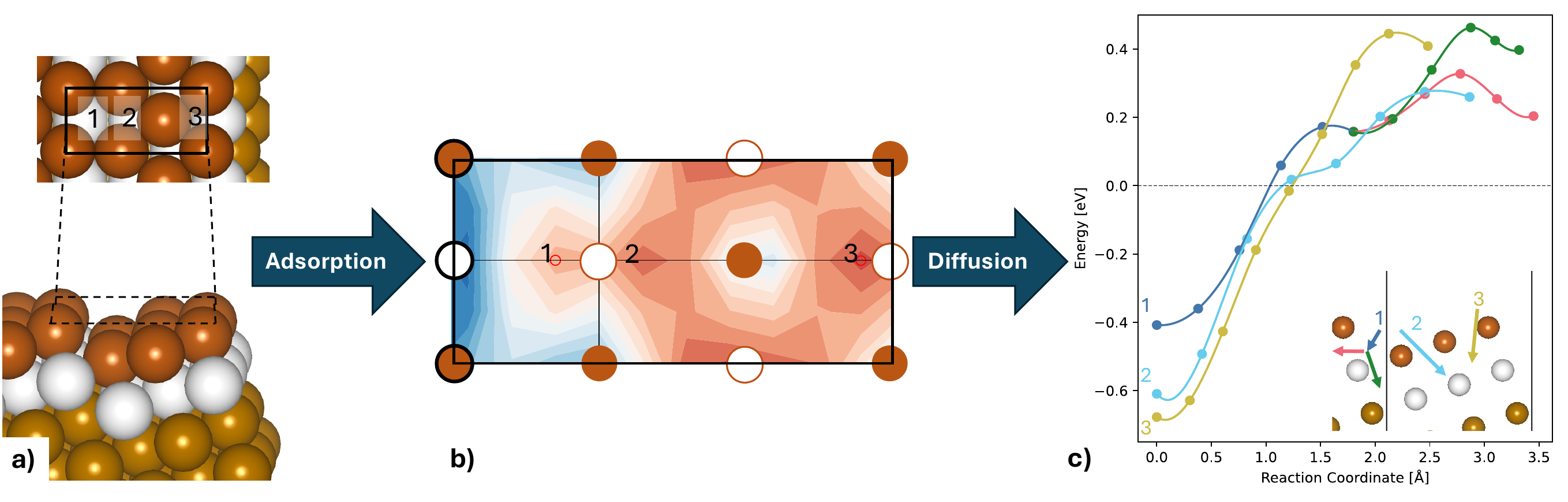}
    \caption{Exemplary representation of our workflow: \textbf{a)} shows the (210) surface unit cell with its top-view projection. \textbf{b)} highlights the potential energy surface for single hydrogen adsorption. Different areas of darkred correspond to local adsorption minima as indicated by the different numbers. Using the different adsorption sites on the surface as starting points for hydrogen diffusion, we calculated the diffusion pathways and energetic profiles for diffusion into the subsurface as shown in \textbf{c)}.}
    \label{fig:ads_pes_scheme}
\end{figure}

As a first step, hydrogen adsorption on top of the different investigated surfaces was probed. To this end, we calculated the potential energy surface (PES) of atomic hydrogen on top of the different iron surfaces to identify areas of local adsorption minima, see Fig.~\ref{fig:ads_pes_scheme}~a,b. Afterwards, the adsorption energies at identified minima were calculated by fully relaxing the hydrogen atom.
The resulting adsorption energies for each surface are plot as different datapoints in Fig.~\ref{fig:E_ads}. 

\subsection{Geometry-Energy correlation}

\begin{figure}
    \centering
    \includegraphics[width=0.6\linewidth]{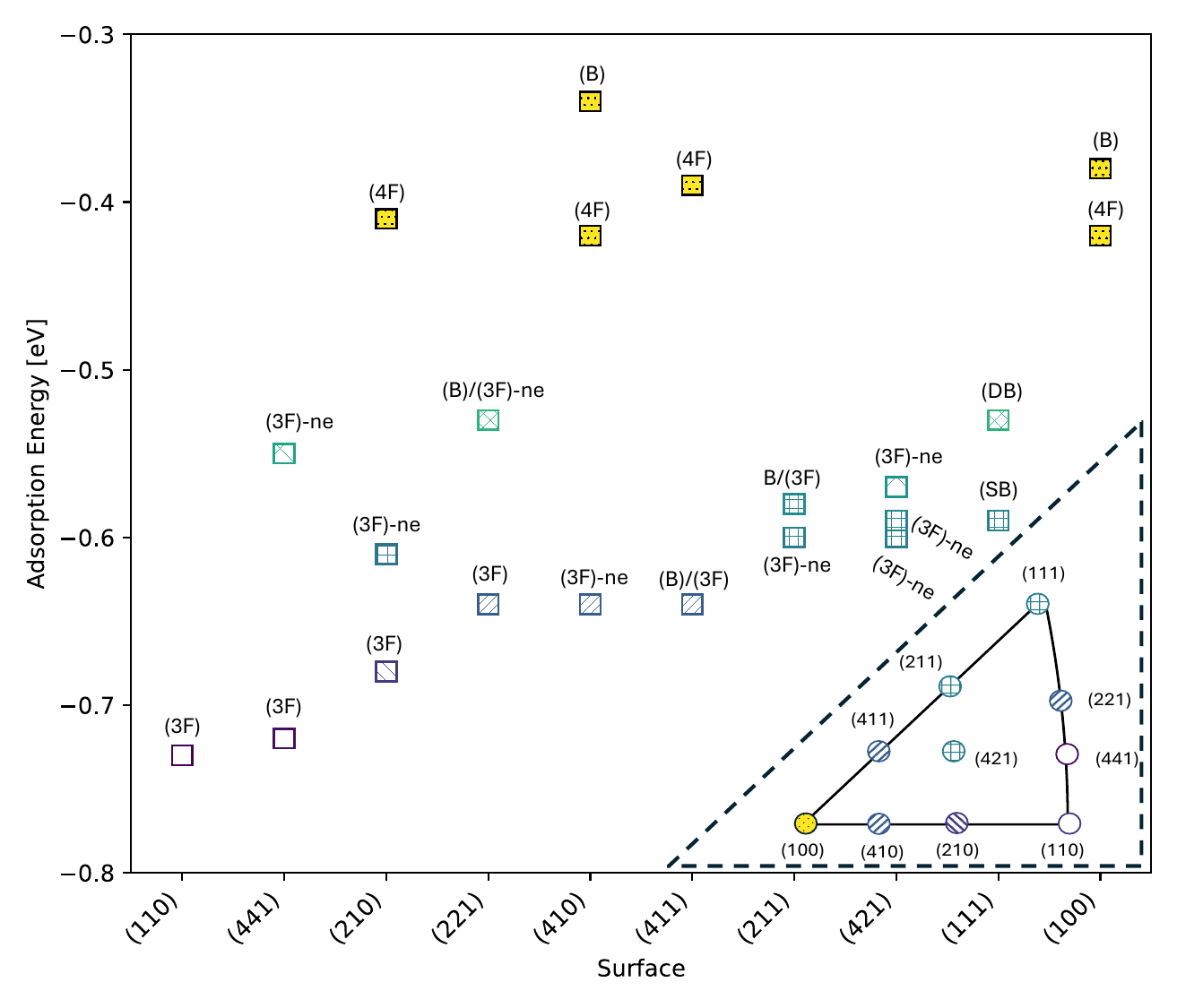}
    \caption{For each investigated surface, hydrogen adsorption energies at all local adsorption minima are plot. The surfaces are sorted in increasing order based on their lowest (i.e., strongest) adsorption energy. The labels and texture of the data points corresponds to their respective adsorption geoemetry and energy, see text. Inset: Representation of the globally strongest adsorption energy of the investigated surfaces in the stereographic projection.}
    
    \label{fig:E_ads}
\end{figure}

First, it can be seen in Fig.~\ref{fig:E_ads} that the higher index surfaces have generally more distinct local adsorption minima than their low-index equivalents, owing to their reduced symmetry.
Secondly, adsorption on the (110) surface is globally found strongest, while hydrogen adsorption on the (100) surface is among the weakest. Although we only studied a limited number of surface orientations, they are well spread among different geometries as discussed in Sec.~\ref{sec:intro}, pointing out the special role of the (110) and (100) surface as an upper and lower limit for hydrogen adsorption energies.
To study the different surface adsorption sites w.r.t. their corresponding adsorption energies, the surface orientations are sorted according to their energetically most favorable adsorption minimum in increasing order. The adsorption energies are color-coded from $-0.73$~eV (purple) over $-0.59$~eV (teal) to $-0.42$~eV (yellow). 
The labels of the data points correspond to the respective adsorption geometry: (4F), (3F), (B), (SB), and (DB) stand for fourfold, threefold, bridge, shallow bridge, and deep bridge site. The appendix \say{ne} indicates that the adsorption site is located near the step edge of a vicinal plane resulting in at least Fe atom has an increased number of nearest neighbors, compared to the flat low-index surface.
Comparing the globally lowest adsorption energies for every surface, a gradual increase from $-0.73$~eV (110) up to $-0.53$~eV (111) is found. The corresponding adsorption geometries are exclusively threefold or mixed threefold/bridge sites, resembling the adsorption geometry on the (110) and (111) surfaces.
On the other hand, the (210), (410), and (411) surfaces show a large spread between their different local adsorption minima exhibiting secondary adsorption sites with a fourfold geometry, which resembles the adsorption minimum of the (100) surface ($-0.42$~eV). 

What is the common trend in these data?
In our recent review paper \cite{Meier2024}, we suggested a correlation between the (global) adsorption energy of hydrogen on a surface and the geometry of that surface represented by its stereographic projection.
When plotting the lowest-energy datapoints in a stereographic projection, we see that the proposed trend spontaneously emerges (see Fig.~\ref{fig:E_ads}, inset): surfaces with similar global adsorption energies are located in close vicinity to each other, as indicated by a continuous coloring scheme.
An exception to this poses the (100)-surface, as geometrically similar surfaces with (100) vicinal planes (such as (411), or (410)) possess an energetically lower global adsorption minimum with a threefold geometry that is located at the step edge region connecting two low-index terraces (see Fig.~\ref{fig:PES_vic_comp}). This renders the (100)-like adsorption site only a secondary adsorption minimum for these surfaces.

In a next step, we compare the electronic structures associated with hydrogen adsorption on different surfaces with a comparable adsorption energy.
This way, we will get a better understanding of the underlying physical reasons leading to the observed trend.

\subsection{Electronic Structure of Hydrogen Adsorption on (100), (110), and (111)}

\begin{figure}
    \centering
    \includegraphics[width=1\linewidth]{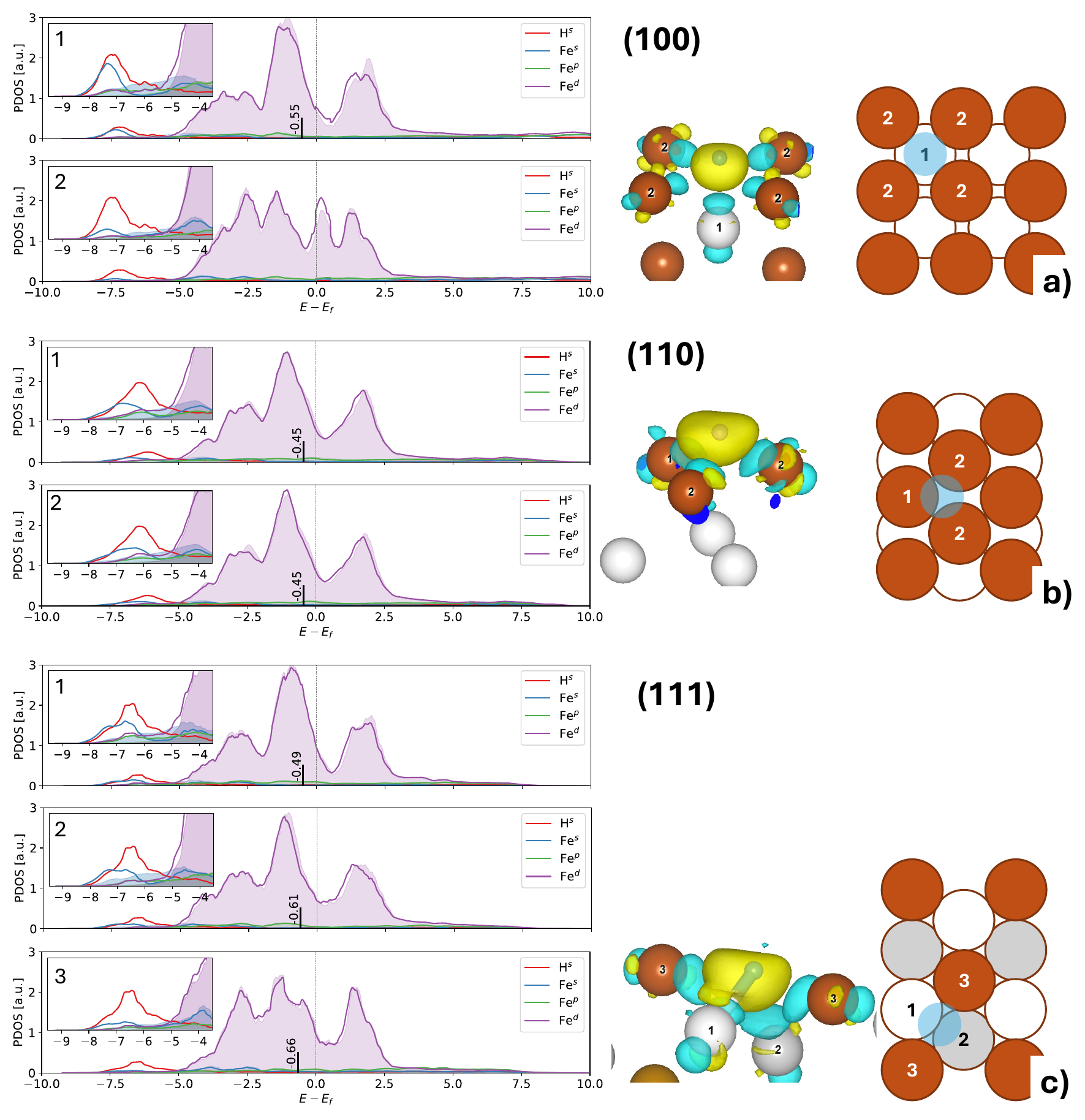}
    \caption{
    Partial density of states (PDOS) of the adsorbed hydrogen atom and the nearest surface iron atom at different local minima on \textbf{a)} the fourfold (4f)-(100) surface site, \textbf{b)} the threefold (3f)-(110) surface site, and \textbf{c)} the shallow bridge (DB)-(111) surface site. The filled curve transparent plots correspond to the PDOS plots before hydrogen adsorption and the solid lines indicate the PDOS distribution after hydrogen adsorption. The d-band center of the pristine surface iron atoms is indicated by a tickmark. The charge density difference plots for hydrogen adsorption are given for an isosurface value of 0.003~eV/Å with yellow and blue areas indicating a charge accumulation and depletion, respectively. We used the VASPKIT code for post-processing of the VASP calculated electronic structure data \cite{vaspkit}.
    The surface iron atoms involved in the bonding are indicated in the right panel with the labels corresponding to the PDOS plots in the left panel.}
    \label{fig:PDOS_low_index_surfaces}
\end{figure}

To set a ground for the subsequent discussion in Sec.~\ref{sec:ES_high_index}, we first discuss the properties of the electronic bonding characteristics of the three low-index surfaces as a reference.
In Fig.~\ref{fig:PDOS_low_index_surfaces}, the global minimum bonding site as well as a charge density difference plot are indicated for each surface.
We highlight that 
\begin{itemize}
    \item Bonding on the fourfold (4F)-\textbf{(100)} site is primarily mediated through the s- electrons of the four adjacent surface atoms. The second layer iron atom weakly binds with its p- and d-electrons to the H-1s orbital, comparable to hydrogen adsorption on-top of a surface atom.
    \item The \textbf{(110)} surface exhibits only one stable hydrogen adsorption site located at a (quasi-) threefold (3F) position. Hydrogen adsorption is decisively determined by an overlap with the d-electrons of the three adjacent surface atoms. Within the framework of d-band theory, the relative position of the d-band center gives a qualitative estimate of how strongly the respective iron atom's d-electrons contribute to the final adsorption energy.
    As the d-band center of the three (110)-surface atoms is located closer to the fermi level than the d-band center of the (100)-subsurface atom, we can qualitatively explain the stronger adsorption energy on the (110) surface.
    \item Due to the open geometry of the \textbf{(111)}-surface, an adsorbing hydrogen atom binds mainly to two adjacent atoms in a bridge-site position. There are two different adsorption minima available. At the deep bridge (DB)-site between the second and third layer atom, we find an overlap with s-, p- and d- electrons. The d-band center is located deeper below the fermi level than at the (110) surface, resulting in an adsorption energy that is weaker than adsorption on the (110) surface, but stronger than on the (100) surface.
\end{itemize}

A detailed discussion covering each of the low-index surfaces for the technically interested reader can be found below. The global story continues at Sec.~\ref{sec:ES_high_index}

\subsubsection{(100)} 
Although the (100)-surface has two local adsorption minima (see Fig.~\ref{fig:E_ads}), the energetically higher bridge site only constitutes a very shallow minimum connecting the diffusion pathway between two adjacent fourfold (4F)-sites, which act as starting points for subsequent diffusion into the subsurface. 
It can be seen that in case of the fourfold (4F)-(100) site (a), the four nearest neighbor surface iron atoms (indicated in brown), as well as the underlying second-layer iron atom (in white) are involved in the bonding process, as indicated by the charge density difference plots signaling a charge transfer from the iron atoms to the adsorbed hydrogen atom.
Since the four surface atoms are at geometrically equivalent positions, it is sufficient to look at one of these atoms (label 2) together with the second surface layer atom (label 1). The electronic density of states plots indicate that the hydrogen electron peaks at an energy around 7.5~eV below the fermi level. 
Comparing the shift (difference) between the relative position of the electronic states of the surface iron atoms prior and after hydrogen adsorption to overlap with the H-1s state, we can deduce on the involved bonding mechanism. In case of the surface iron atoms (2), a shift of their s-electrons to overlap with the hydrogen orbital at $~-7.5$~eV is observed (blue line), while no significant d- and p-electron shifts are visible (green and purple lines).
For the second layer iron atom (1), next to an s-electron shift, also less pronounced p- and d-electron peaks around 7~eV below the Fermi level emerge that are not visible for the surface iron atoms.
This is in line with observations of hydrogen adsorption on top of a single iron surface atom, which involves the more directional d-orbitals with an orientation along the z-direction, as discussed in \cite{Jiang2003}. On the other hand, the s-orbitals are less directional corresponding mainly to the bonding of the near-surface site in the center of the fourfold-coordinated site (see discussion in \cite{Meier2024}).
This means that bonding to the (4F)-(100) site is mainly mediated via s-electrons of the four adjacent surface iron atoms, while the second layer iron atom's s-, p-, and d-electrons contribute to the bonding, in line with previous results \cite{Boda2019}.

\subsubsection{(110)}\label{sec:ES_110}
For H adsorption at a threefold (3F)-(110) site, only the three nearest surface atoms show a visible charge transfer to the hydrogen atom (b). Due to the D$_2$ symmetry of the surface, they do not form an equilateral, but rather an isosceles triangle, which leads to a geometrically slight difference between the summit of the triangle (atom 1) and its base (atom 2). Examining the PDOS graphs, the electronic density of states is however almost identical. We find that for all atoms next to a visible s- and p-electron shift, a pronounced d-band shift to overlap with the hydrogen orbital at an energy about $6$~eV below the fermi level. This means that the hydrogen bonding at the (3F)-(110) site is significantly determined also by the Fe-3d electrons in contrast to adsorption on the (4F)-(100) site and in agreement with previous studies \cite{Jiang2003, Li2020, Chohan2016, Li2019_2}.
Due to the decisive role of the d-electrons, it becomes interesting to look at the d-band center of the surface iron atoms prior to hydrogen adsorption. For the (110)-surface atoms the d-band center is found to be located 0.45~eV below the Fermi level. This is 0.1~eV higher than the relative position of d-band center of atom (1) at the (4f)-(100) site, while the overall shape of both d-bands are similar.
This is expected as a deeper d-band center is usually associated with a stronger bound atom, which is the case for the second-layer atom in a), which has a complete first coordination shell, compared to the surface atoms in b).
In the framework of d-band theory \cite{Bligaard2007, Wei2020}, which is commonly applied for describing catalytic properties of metal surfaces, this implies that the bonding strength of the d-electrons is weaker for the (100) atom than for the (110)-atoms with the higher relative d-band center. 
This constitutes at the same time a qualitative explanation to why H adsorption on the (110) surface is much stronger than on the (100) surface: The d-band electrons of all nearest-neighbor surface iron atoms contribute to hydrogen bonding with a d-band center located close to the Fermi level.

\subsubsection{(111)}
For the (111)-surface, we examine the adsorption of hydrogen at the deep bridge (DB) site with an adsorption energy of -0.53~eV. While it has a higher adsorption energy than the shallow bridge (SB) site (-0.57~eV), hydrogen first diffuses from the (SB) to the (DB) site before entering the subsurface, rendering the (DB) site the starting site the starting point for surface-to-subsurface hydrogen diffusion.
Bonding at the (DB)-(111) surface site is primarily mediated by the two involved bridge atoms, denoted as atom~(1) in the second layer and atom~(2) in the third surface layer.
Even though the charge density difference plots suggest also a contribution of the surface-layer atom~(3), it can be seen that the overlap of its atomic orbitals is minimal compared to the other two atoms, indicating only a minor contribution to adsorption.
The relatively lower number of bonding partners at this site arises from the more open surface structure, which leads to a larger lateral distance between atoms in the same layer.
Atom~(1) shows a strong overlap of its s-orbital and a moderate overlap of its p- and d-band with the H-1s electronic band. Atom~(2) misses the p-band interaction, as a result for the asymmetry between both atoms. The d-band shapes of both atoms are similar, resembling the d-band shapes typically observed for atoms on the (110) surface. 
With both bridge atoms exhibiting a partial shift in their d-bands following hydrogen adsorption, the position of the d-band center relative to the Fermi level provides further insight into the bonding characteristics. The d-band center of atom~(1) is located 0.49 eV below the Fermi level, a position comparable to that of the (3f)-(110) surface atoms. In contrast, the d-band center of atom 2 is positioned deeper below the Fermi level, suggesting a reduced contribution to bond strength.
Overall, bonding at the (DB)-(111) site is mediated by the s- and d-electrons of the two bridge atoms. The asymmetry in the electronic structure, arising from their positions in different surface layers, results in the second-layer atom~(1) contributing more significantly to the bonding energy than the third-layer atom~(2) via its d-orbital. Consequently, the total adsorption energy of hydrogen at the (DB)-(111) site is higher than the adsorption energy at the (3F)-(110) site, but lower than adsorption at the (4F)-(100) site.

\subsection{Electronic Structure Comparison of higher index surfaces}\label{sec:ES_high_index}
After studying the electronic properties of hydrogen adsorption on the three low-index surfaces as reference points, we now compare them with the electronic bonding characteristics of hydrogen adsorption on other surfaces at geometrically similar adsorption sites, which exhibit a similar adsorption energy (see Fig.~\ref{fig:E_ads}). 
This means in the case of vicinal surfaces that we compare hydrogen adsorption near their step-edge, where locally the adsorption geometry on the vicinal planes resembles the adsorption scenario of the low-index surfaces, but with the step edge as a potential influence factor nearby.

The main results are summarized below with a more in-detail discussion in Sec.~\ref{sec:100_111-vicinal_planes} and \ref{sec:(110)-vicinal_planes}. A reader less interested in those details can skip these sections and move on to Sec.~\ref{sec:spatial_extent} after the summary.

\begin{itemize}
    \item We first compare hydrogen adsorption on the (100)-surface and near the step edge of the (410)-surface, exhibiting a planar structure of (100)-vicinal planes. 
    We find that the adsorption scenario remains virtually unchanged with only the shape of the electronic d-band of the near-edge surface atoms altered. While our electronic structure comparison is limited to a single vicinal surface, it is reasonable to assume that this is a more general trend due to the larger number of surfaces that possess a secondary (100)-like adsorption minimum within a narrow interval around the (4F)-(100) adsorption energy (see the yellow points in Fig.~\ref{fig:E_ads}).
    
    \item Due to the open structure of the (111)-surface, there are no vicinal surfaces in our test set with an extended (111)-plane. The (221)-surface is a minimal example, which only allows for a limited comparison. As a consequence, the hydrogen atom tends to bind in a threefold geometry, which is a closer resemblance of the (3F)-(110) site with the edge atom as the third bonding partner. This behavior is confirmed by a stronger overlap with the iron d-electrons, although the adsorption energy is higher than on the (110) surface, due to deeper d-band center locations.
    
    \item We compare hydrogen adsorption on the (110)-surface with adsorption near the step edge of the (441)-vicinal surface and near an adatom on the (100)-surface. These configurations locally resemble the threefold (3F)-(110) adsorption geometry with only the three adjacent iron atoms as bonding partners. Nevertheless, we find a strong difference with the corresponding adsorption energy near the adatom and close to the step-edge of the (441)-vicinal surface moderately increased compared to the (110) adsorption energy. This is because surface iron atoms near the step edge have a higher coordination number than equivalent atoms on a flat surface, rendering them stronger bound. This leads to a lowering of their d-band center, which indirectly increases the hydrogen adsorption energy to these atoms. The increased adsorption energy is then similar to the adsorption energy on the (111) and (221)-surface. 
\end{itemize}

\begin{figure}
    \centering
    \includegraphics[width=1\linewidth]{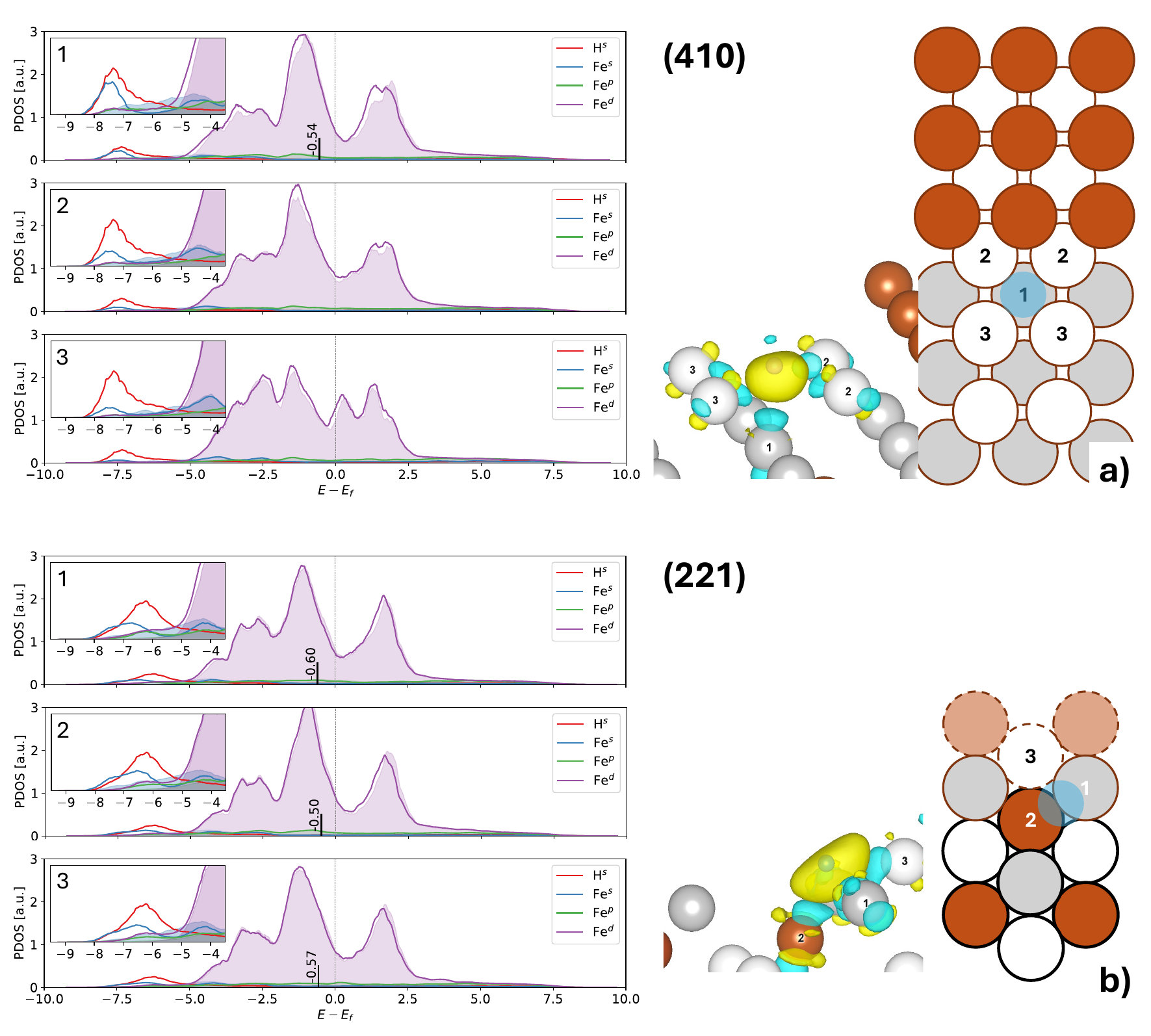}
    \caption{Partial density of states (PDOS) of the adsorbed hydrogen atom and the nearest surface iron atom at different local minima on \textbf{a)} the (410) surface resembling the local geometry of the (100) surface, and \textbf{b)} a bridge site on the (221) surface, resembling the (DB)-(111) surface site (see Fig.~\ref{fig:E_ads}, circles). The filled curve transparent plots correspond to the PDOS plots before hydrogen adsorption and the solid lines indicate the PDOS distribution after hydrogen adsorption. The d-band center of the pristine surface iron atoms is indicated by a tickmark. The charge density difference plots for hydrogen adsorption are given for an isosurface value of 0.003~eV/Å with yellow and blue areas indicating a charge accumulation and depletion, respectively. We used the VASPKIT code for post-processing of the VASP calculated electronic structure data \cite{vaspkit}.
    The surface iron atoms involved in the bonding are indicated in the right panel with the labels corresponding to the PDOS plots in the left panel.}
    \label{fig:PDOS_410_211}
\end{figure}

\subsubsection{Adsorption on (100)- and (111)-vicinal planes}\label{sec:100_111-vicinal_planes}

In Fig.~\ref{fig:PDOS_410_211}~a, we study adsorption on the (410) surface as a vicinal surface with (100) planes. 
According to our hypothesis, we expect for hydrogen adsorption on these planes to resemble the same bonding picture as obtained for hydrogen adsorption on the (4f)-(100) site (Fig.~\ref{fig:PDOS_low_index_surfaces}~a).
Looking at the charge density difference plot, we can see that again the same nearest neighbor atoms are involved in the bonding: The underlying second layer atom (1), and the four nearest neighbor surface atoms split up in two groups (2) and (3) due to the nearby step edge reducing the symmetry compared to the pristine (100) case.
Comparing the PDOS plots with the ones in Fig.~\ref{fig:PDOS_low_index_surfaces}, virtually the same picture emerges: For atom~(1), we have an additional d-band shift with with its d-band center at the same position relative to the Fermi level as it was for the (100) surface. The only difference arises from the d-band shape of (2) compared to (3) and the surface atoms of the (100) surface. This can be explained by the influence of the nearby step-edge, which increases the surface iron atom's coordination number to resemble locally the (110)-surface geometry. However,this does not visibly affect the electronic structure around the H-1s peak, which results in the same adsorption energy as on the (100) surface.

Next, we compare hydrogen adsorption on (221)-(B) bridge site (Fig.~\ref{fig:PDOS_410_211}~b) with adsorption on the (111)-(DB) site (Fig.~\ref{fig:PDOS_low_index_surfaces}~b). Although less clear than for the vicinal surfaces with (110) and (100) planes, the (221) surface can be seen as a minimal example for a vicinal surface with (111)-planes if oriented accordingly, as indicated in schematic sketch of Fig.~\ref{fig:PDOS_low_index_surfaces}~b. In contrast to the (111)-(DB) site, hydrogen is adsorbed close to the step edge marked by the atom~(3). Thus, atom~(3), which corresponds to the surface-layer atom~(3) in Fig.~\ref{fig:PDOS_low_index_surfaces}~c is only present on one side of the hydrogen atom, leading to a slightly skewed symmetry and effectively makes renders the (221)-(B) site an intermediate version between shallow and deep bridge with its adsorption energy at the (221)-(B) site coincides withe the (111)-(DB) adsorption energy of -0.53~eV.
In contrast to the (111)-(DB) site adsorption scenario, we observe at the (221)-(B) site a non-negligible contribution of atom~(3) expressed by an overlap of its s-, p-, and d-orbitals with the H-1s orbital. Moreover, there is no visible difference in the shape of the d-bands and the involved orbitals between the three iron atoms, resembling more a (110)-(3F) bonding situation. The only difference arises from the relative position of the d-band center, which indicates a stronger contribution of atom~(2), which effectively resembles the near-edge (110)-like situation discussed below (Sec.~\ref{sec:(110)-vicinal_planes}).
This might be an indication that if both, a (111)-like bridge site and a threefold site resembling the (110)-(3F) situation are locally present, we find a preference towards a threefold-like situation with more bonding partners over a (111)-like bonding site.

\begin{figure}
    \centering
    \includegraphics[width=1\linewidth]{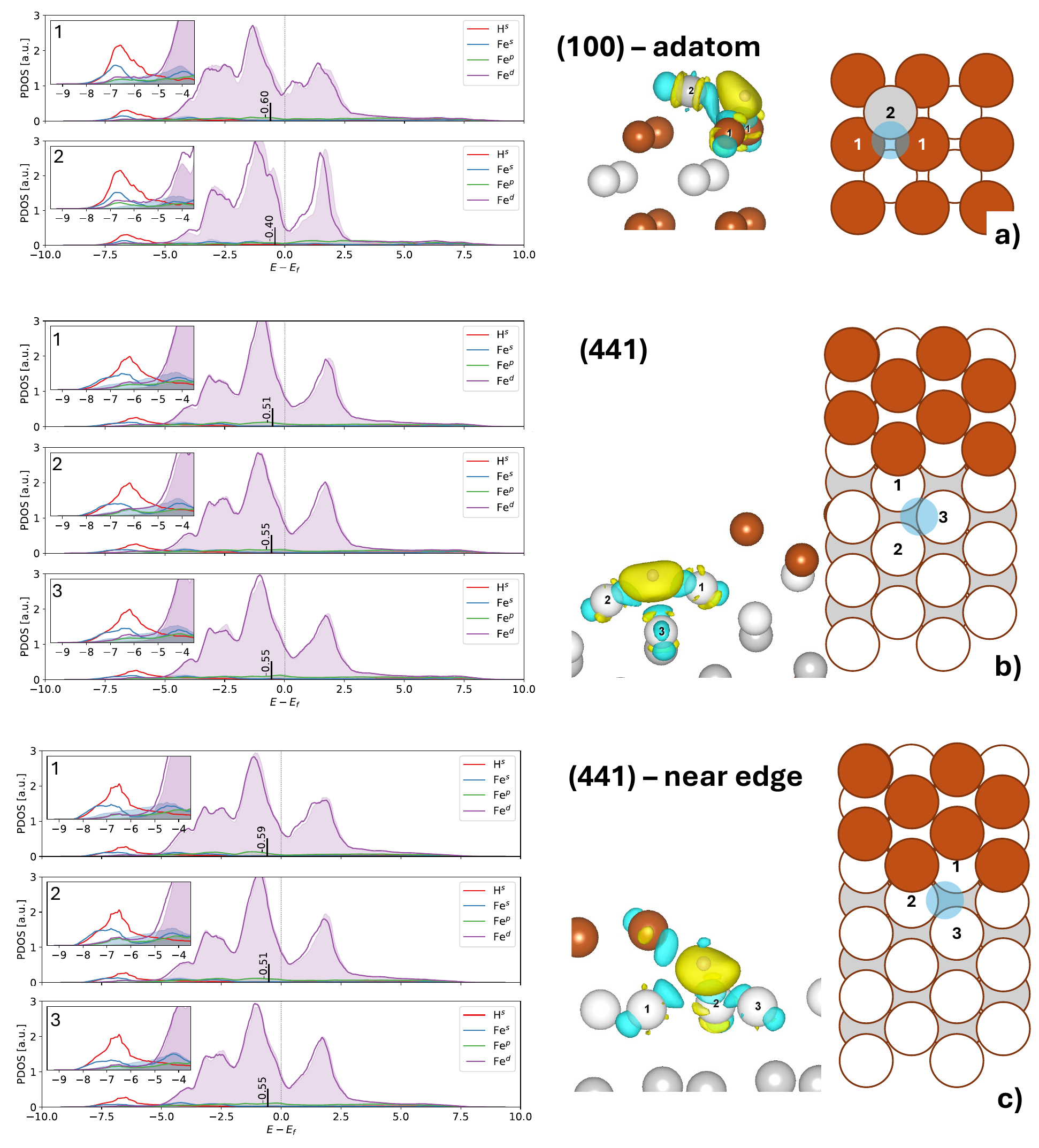}
    \caption{Partial density of states (PDOS) of the adsorbed hydrogen atom and the nearest surface iron atom at different local minima on different surface sites that resemble the local geometry of the (3f)-(110) site: \textbf{a)} An adatom on the (100) surface, \textbf{b)} (441) surface near the step edge, \textbf{c)} (441) surface further away from the edge (see Fig.~\ref{fig:E_ads}, circles). The filled curve transparent plots correspond to the PDOS plots before hydrogen adsorption and the solid lines indicate the PDOS distribution after hydrogen adsorption. The d-band center of the pristine surface iron atoms is indicated by a tickmark. The charge density difference plots for hydrogen adsorption are given for an isosurface value of 0.003~eV/Å with yellow and blue areas indicating a charge accumulation and depletion, respectively. We used the VASPKIT code for post-processing of the VASP calculated electronic structure data \cite{vaspkit}.
    The surface iron atoms involved in the bonding are indicated in the right panel with the labels corresponding to the PDOS plots in the left panel.}
    \label{fig:PDOS_110_plane}
\end{figure}

\subsubsection{Adsorption on (110)-vicinal planes}\label{sec:(110)-vicinal_planes}

Last, we compare different local environments that resemble the (3F)-(110) site in Fig.~\ref{fig:PDOS_110_plane}: An adatom on the (100) surface, and two sites on the vicinal (110)-plane of the (441) surface, one close to the edge, the other further away.
According to the hypothesis that local geometry defines adsorption, we would expect them to exhibit about the same adsorption energy as on the pristine (110) surface. Interestingly, this is only the case for the adsorption site further away from the edge (b), while close to the edge of the (441) surface and near the (100)-adatom, the adsorption energies are found moderately reduced to about $-0.55$~eV, compared to $-0.72$~eV on the pristine (110)-surface.
What causes this reduction of adsorption energy close to the step edge of (110)-vicinal surface planes?
It can be seen that each of the surface iron atoms involved in the bonding undergoes a hybridization not only of its s- and p-orbitals, but also of its d-orbitals upon hydrogen adsorption, in alignment with our discussion on the (110) surface in Sec.~\ref{sec:ES_110}. 
Therefore, the context of the d-band theory can be employed to explain the observed hydrogen adsorption behavior \cite{Bligaard2007, Wei2020}: The d-band center of the surface iron atoms with more nearest-neighbour bonding partners (as it is the case closer to the step edge or adjacent to an adatom) is located further below the Fermi level ($-0.60$ and $-0.59$~eV) than the d-band center of surface iron atoms with a lower coordination number ($-0.45$ to $-0.55$~eV). This is a consequence of the near-edge surface atoms having a higher coordination number, thus being stronger bound within the surface structure. At the same time, the relative location of the d-band center w.r.t. the Fermi level is a direct measurement of the bonding strength: Upon hybridization, the d-orbital is split up into a bonding and an anti-bonding part. The bonding orbital is located deeper below the Fermi level and completely filled, while the  anti-bonding orbital is centered around a higher energy and only partially filled. If the d-band center is located at a higher energy and thus, closer to the Fermi-level, the energy of the anti-bonding orbital will also be increased, rendering it less filled, thus, increasing the adsorption energy at that site.
Therefore, the reduced adsorption energy for (a) and (c) can be explained by a lower d-band center location of the involved iron surface atoms as a result of the nearby edge atoms.

\subsection{Vicinal Surfaces: Spatial Extent of the step-edge influence}\label{sec:spatial_extent}
In the previous section, we discussed that hydrogen adsorption near a step edge of (110)-vicinal surfaces is weaker than further away from the edge. At the same time, it can be expected that within a certain distance, the step region will also influence the potential energy surface of other vicinal surfaces. Therefore, the question arises, up to what spatial extend the step edge affects hydrogen adsorption on vicinal planes.

\begin{figure}
    \centering
    \includegraphics[width=1.0\linewidth]{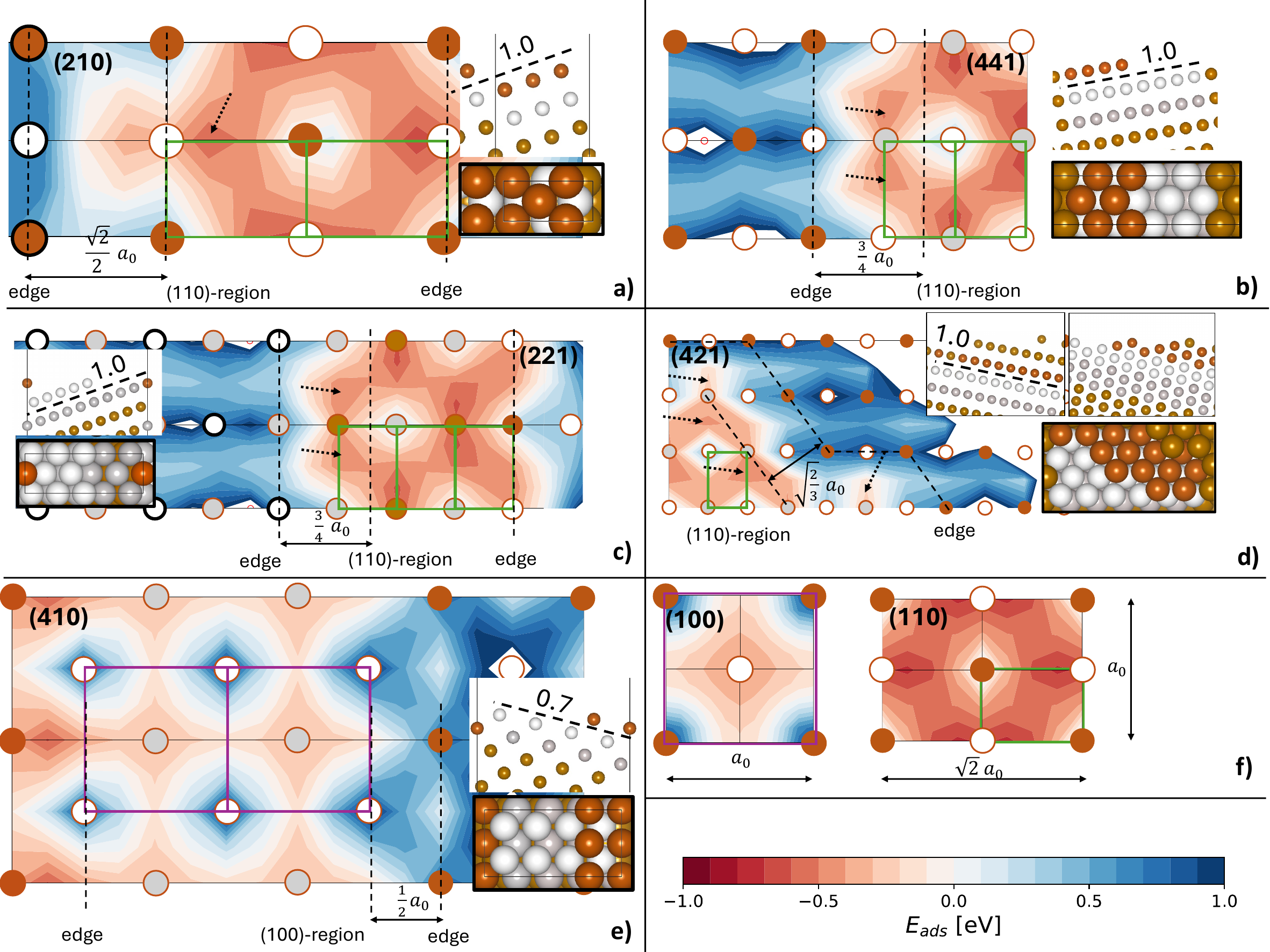}
    \caption{Potential energy surface (PES) of hydrogen at a fixed height above the low index planes of the investigated vicinal surfaces as indicated in the insets of each figure (in Å): \textbf{a)} (210), \textbf{b)} (441), \textbf{c)} (221), \textbf{d)} (421), \textbf{e)} (410). The PESs of (100) and (110) are provided for comparison in \textbf{f)}.
    }
    \label{fig:PES_vic_comp}
\end{figure}

The spatial extent of the step interrupting the terrace-structure can be gauged by comparing the PESs for hydrogen adsorption at a fixed height parallel to the vicinal surface planes (see Fig.~\ref{fig:PES_vic_comp}). It can be seen that the step induces a transition zone, where the adsorption profile is changed compared to the low-index surfaces. This zone has a lateral extension parallel to the low-index planes of approximately $0.5~a_0 \approx 1.4$~Å and $0.75~a_0 \approx 2$~Å for (100)- and (110)- vicinal surfaces, respectively, and is mainly present on the lower terrace next to a step, while the region approaching the edge from the upper terrace remains largely unaffected (see Fig.~\ref{fig:PES_vic_comp} a, c, e). 
In agreement with our discussion in Sec.~\ref{sec:(110)-vicinal_planes}, adsorption energies are found increased (i.e., weakened) near the step edge of (110)-vicinal surfaces (see Fig.~\ref{fig:PES_vic_comp} a-d, dashed arrows).

\section{Diffusion}
The crucial step in determining the susceptibility of a metallic surface to hydrogen uptake occurs when the adsorbed hydrogen diffuses from the surface into the subsurface region of the material \cite{Meier2024, Jiang2004}. During this step of the diffusion process, the diffusion behaviour deviates most from the isotropic bulk behaviour and the bonding character changes from an exothermic bonding state on the surface to an endothermic state in the material, for which the subsurface acts as the transition zone.

\begin{figure}
    \centering
    \includegraphics[width=1\linewidth]{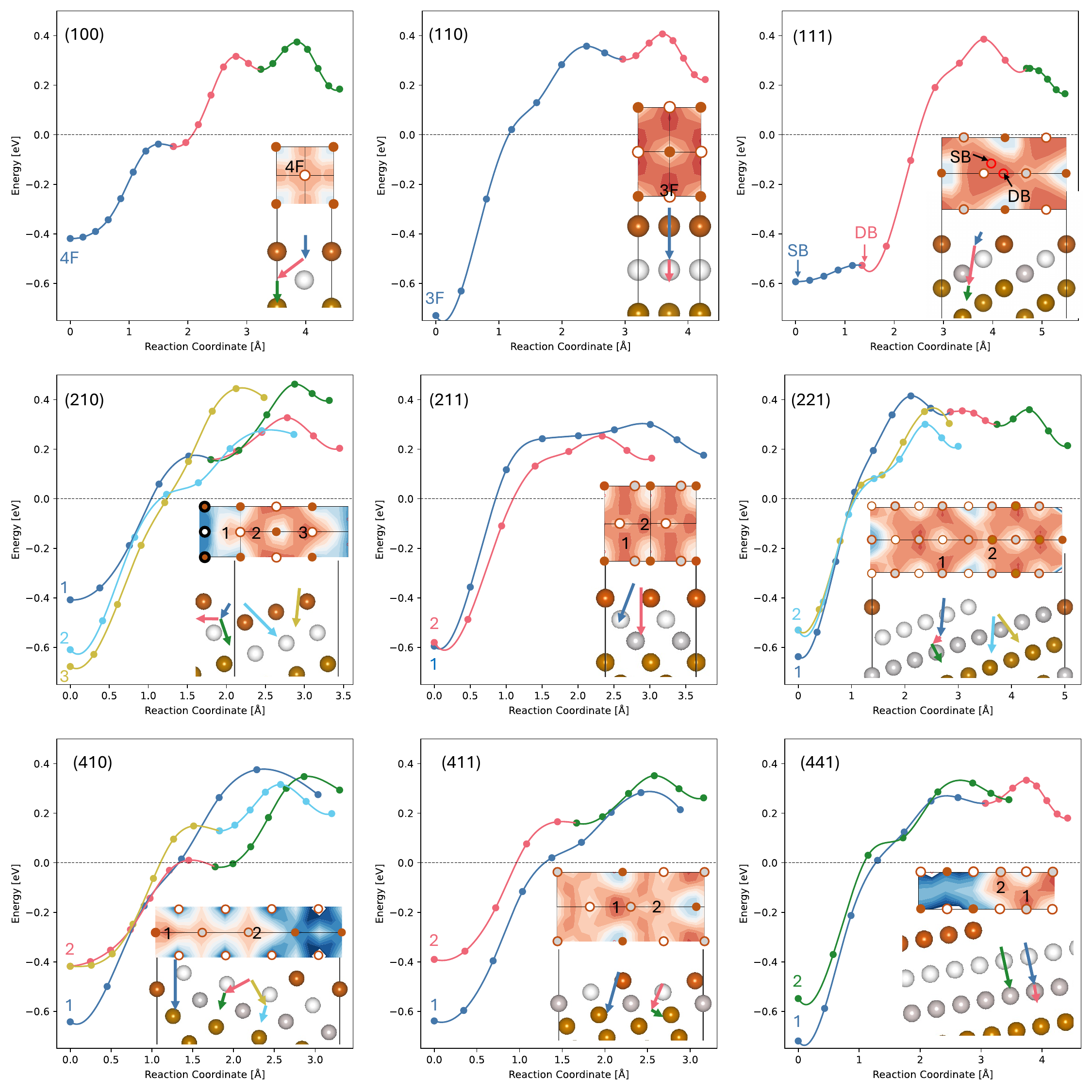}
    \caption{NEB-profiles of different possible diffusion pathways of atomic hydrogen through iron surfaces. The corresponding paths are indicated with matching colors in the corresponding sub-figures. The local adsorption minima on the surfaces chosen as starting points for the NEB calculations are indicated in the corresponding PES plots. The zero-point of the energy scale is set according to Eq.~\ref{eq:adsorption_energy} as the hydrogen atom in its molecular form far away from the surface.}
    \label{fig:NEB-grid}
\end{figure}

To obtain a comprehensive overview of the role of different surface orientations in hydrogen adsorption, we calculated the surface-to-subsurface diffusion profiles for 9 different surfaces, along different paths. These include the low-index surfaces and six other (vicinal) surfaces. Therefore, the different adsorption minima on top of the surface act as starting points of diffusion paths connecting to adjacent tetrahedral sites in the subsurface. The respective adsorption profiles are plotted in Fig.~\ref{fig:NEB-grid} with the lowest barriers for each surface given in Tab.~\ref{tab:E_diff}.

The NEB-calculated diffusion profiles for the low-index surfaces (100), (110), and (111) are shown in the first row of Fig.~\ref{fig:NEB-grid}
Among these, the (100) surface is the easiest permeable because the surface-to-bulk diffusion is split into two subsequent diffusion steps, each of which requires overcoming an energy barrier of only around 0.4~eV. This is considerably smaller than the energy barriers of 1.1 and 0.9~eV, related to diffusion into the (110) and (111) surfaces, respectively, in qualitative agreement with literature \cite{Meier2024, Jiang2004, Shen2016, Boda2019, Sorescu2005, Li2020, Chohan2016}. Physically speaking, this is because for the (100) surface, the first tetrahedral site below the surface is a stable site with an intermediate energy level that splits the energy difference between adsorption energy on the surface and absorption energy in the bulk in two. For the (110) and (111) surface, this site is not stable, meaning that a hydrogen atom placed there would diffuse back to the surface \cite{Jiang2004, Shen2016}. Therefore, diffusion through these three prototype surfaces can be classified into two scenarios, which will be recognized in more general surfaces later on: \textbf{I)} Diffusion is a two-step process with a stable absorption site in the subsurface that splits the associated barrier approximately halfway, similar to the (100)-behaviour. \textbf{II)} Diffusion from surface to bulk is a one-step process resulting in a higher total diffusion barrier height, resembling the (110) and (111)-behaviour.

Compared to the low-index surfaces, all other surfaces exhibit multiple diffusion paths with different barrier heights, while for the low-index surfaces only one path from surface to subsurface is available owing to their increased symmetry.
Among these high-index surfaces, the (210), (410), and (411)-oriented surfaces exhibit at least one type-\textbf{I} diffusion path with a maximum barrier height between 0.4 and 0.6~eV, while permeation through the other surfaces is predicted to be more difficult with diffusion barriers of 0.8 up to over 1.0~eV.

\begin{figure}
    \centering
    \includegraphics[width=0.75\linewidth]{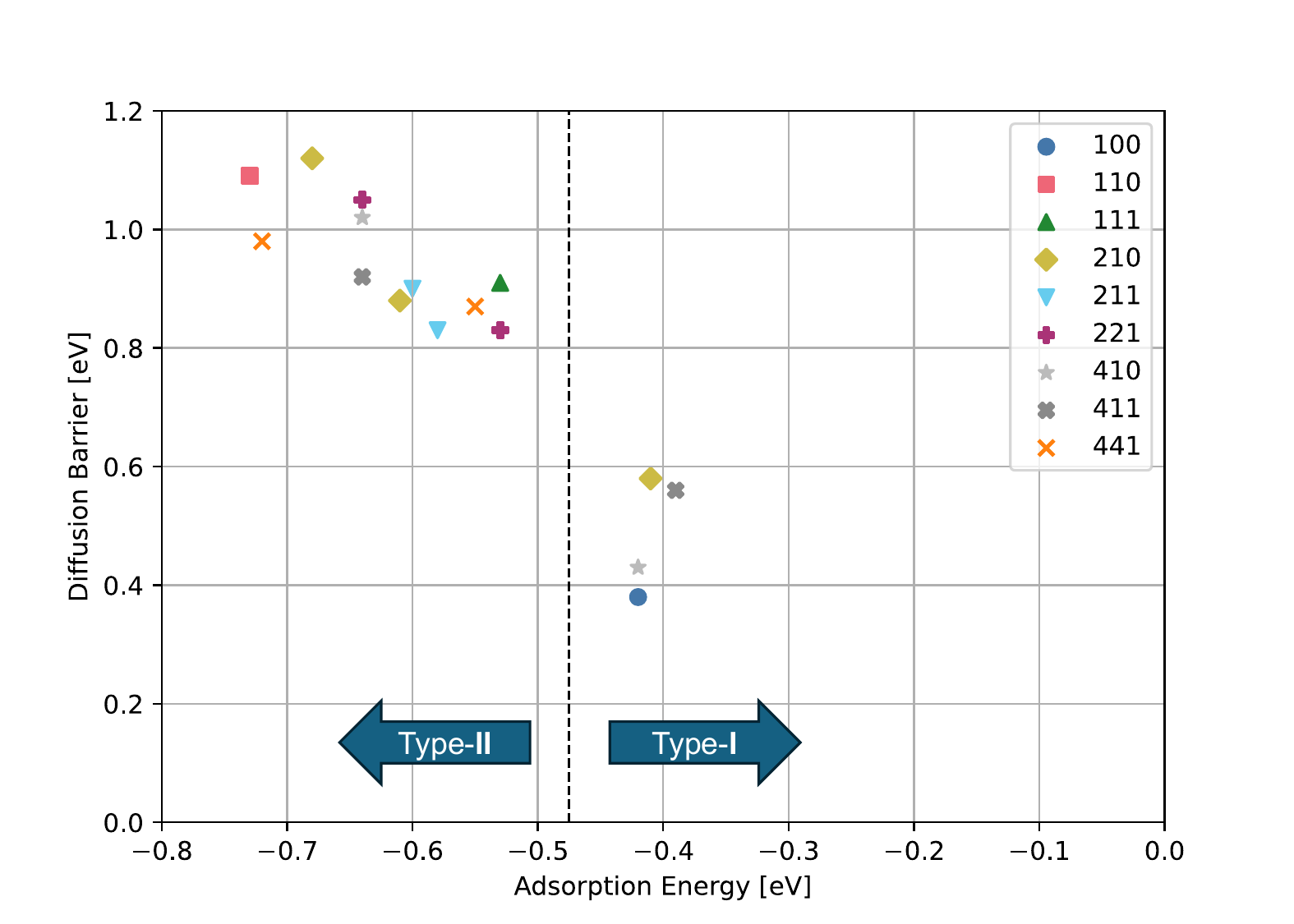}
    \caption{Linear relationship between the adsorption energy at a local minimum on the surface and the corresponding diffusion barrier height into the subsurface using that minimum site as a starting point. Different surfaces are plot using different colors and symbols as indicated in the legend.}
    \label{fig:ads-diff-linear}
\end{figure}

Interestingly, the occurrence of a type-\textbf{I} diffusion barrier seems to be correlated to the existence of a high energy (i.e., (100)-like) adsorption site with an adsorption energy of around -0.4~eV on top of the surface. 
Furthermore, we find that the minimum diffusion barriers for surface-to-subsurface exhibit an almost linear relationship with the corresponding adsorption energies at the starting point on top of the surface with a clear distinction between type-\textbf{I} and type-\textbf{II} regime (see Fig.~\ref{fig:ads-diff-linear}). 
This allows in a first approximation to use the surface adsorption profile to determine, which surfaces are more prone to hydrogen uptake and diffusion: More high-energy adsorption sites indicates an easier hydrogen permeability.

\begin{figure}
    \centering
    \includegraphics[width=1\linewidth]{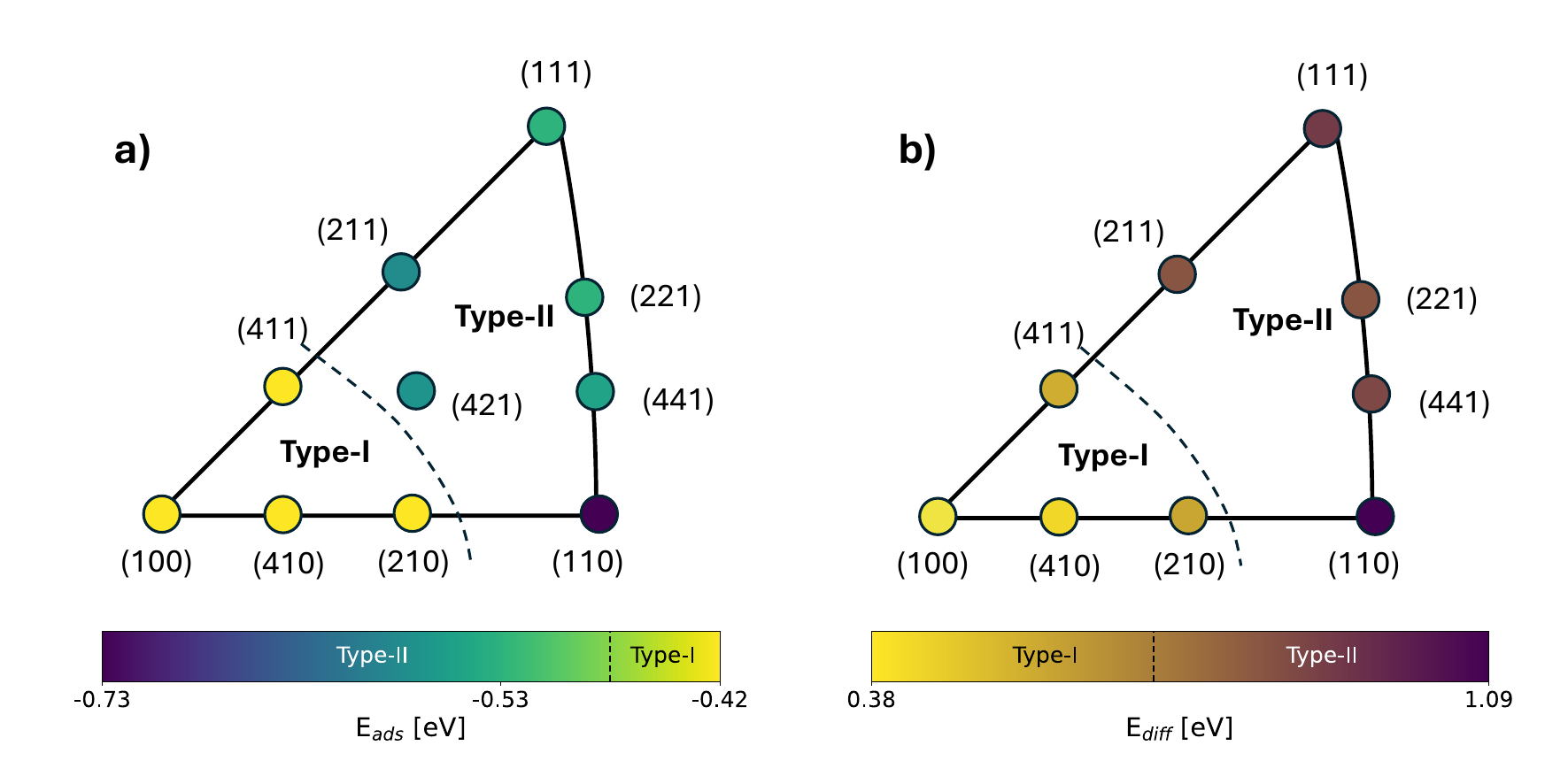}
    \caption{Stereographic projection of all investigated surface orientations. \textbf{a)} Adsorption energies of the starting point of the lowest-barrier surface-to-subsurface diffusion paths are plot. The colors are interpolated between the corresponding adsorption energies on the low-index surfaces (110) (-0.73~eV), (111) (-0.53~eV), and (100) (-0.42~eV). \textbf{b)} Diffusion barrier heights of the easiest surface-to-subsurface diffusion path are plotted with a linear interpolation between the (100) and (110) diffusion barriers.}
    \label{fig:triangle_cmap}
\end{figure}

Next to this energetic criterion, we are further able to determine a geometric criterion, which defines two regions in the stereographic projection, corresponding to areas of type-\textbf{I} and type-\textbf{II} surfaces (see Fig.~\ref{fig:triangle_cmap}): In a), the surfaces are color-coded according to the adsorption energy of the surface site, which is the starting point for the \emph{lowest}-barrier surface-to-subsurface diffusion path. In b), the corresponding diffusion barrier heights are plot. The respective values are given in Tab.~\ref{tab:E_diff}, as well.

\begin{table}
    \centering
    \begin{tabular}{c|cccc|cccccc}
       Surface                & (100) & (410) & (411) & (210) & (221) & (211) & (441) & (111) & (110) & (421)\\
       \hline
       \hline
        E$_{ads}$ [eV]        & -0.42 & -0.42 & -0.39 & -0.41 & -0.53 & -0.58 & -0.55 & -0.53 & -0.73 & -0.57$^*$\\
        E$_{diff}$ [eV] & 0.38 & 0.43 & 0.56 & 0.58 & 0.83 & 0.83 & 0.87 & 0.91 & 1.09 & - \\
    \end{tabular}
    \caption{Overview of the minimum energy barriers ($E_{diff}$) for surface-to-subsurface diffusion with the corresponding adsorption energy ($E_{ads}$) at the starting point for each surface, as plotted in Fig.~\ref{fig:triangle_cmap}}
    \label{tab:E_diff}
\end{table}

Here, some things should be noted: While the diffusion barriers exhibit two regions in the stereographic projection (type-\textbf{I} and type-\textbf{II}), the corresponding adsorption energies are colored three-way, following the adsorption energies of (100), (111), and (110) in yellow, teal, and purple. Therein, the lowest-barrier adsorption energy resembles even for surfaces in close proximity to the (110) surface the adsorption energy at the (111)-DB site, with the (110) surface being the outlier. 
This is because of the increased adsorption energy near the step edge as discussed in Sec.~\ref{sec:(110)-vicinal_planes}, which uniformly corresponds to the lowest-barrier diffusion path and which is present for all (110)-vicinal surfaces. However, this effect is relatively small and only reduces the barrier height compared to starting points further away from the edge only by around 0.1~eV. Furthermore, for larger low-index planes on the vicinal surfaces, the near-step sites will statistically become less relevant for the overall adsorption characteristic, approaching the (110) surface.

Secondly, due to the large cell dimensions, we did not perform diffusion calculations, but only adsorption calculations on the (421)-surface. However, due to its geometric features, it can be classified as a (110)-vicinal surface with skewed terraces, which only allow near step-edge adsorption in line with corresponding adsorption energies, see Fig.~\ref{fig:PES_vic_comp} Thus, it is reasonable to assume similar type-\textbf{II} diffusion barriers as they were observed for comparable vicinal surfaces. Based on this reasoning, the line separating the type-\textbf{I} region from the type-\textbf{II} region was drawn, as shown in Fig.~\ref{fig:triangle_cmap}

\section{Conclusion and Implications}
The adsorption energies, geometries, and electronic structures for single hydrogen adsorption on top of differently oriented monocrystalline bcc-iron surfaces were explored.
It was found that the structural similarity between different surface orientations (including the low-index surfaces and other more general and vicinal surfaces) as expressed by their proximity in the stereographic projection serves as a first approximation to hydrogen adsorption energies. Additionally, the H-adsorption energies near edge-sites of (110)-vicinal surfaces are found moderately reduced, which can be explained in the framework of the d-band center theory.

In a second part, the static diffusion barriers from the surface adsorption sites into the iron subsurface are compared. Therein, a systematic correlation between adsorption site energy and diffusion barrier is found, indicating that the existence of a high-energy adsorption site correlates to a low-barrier diffusion path. As the occurrence of a high-energy adsorption site is linked to the surface's proximity to the (100) low-index surface, we predict surfaces beyond a certain (100)-proximity to be more suitable to prevent hydrogen permeation.
Based on this criterion, we mapped the space of different surface orientations to define a region that marks surface orientations through which it is harder for hydrogen to penetrate into the material.
The proposed map can serve as a guideline to manufacture more hydrogen-repellent steels by maximizing the fraction of grains in the steel's surface texture that belong to the \say{hard} (110)-like (type-\textbf{II}) region.

Experimentally, this directional preference of hydrogen diffusion was indirectly confirmed for the study of hydrogen-assisted crack growth. Here, the pure metallic surface is exposed to the hydrogen gas, which directly mimics our simulations. It was found, that for for hydrogen-enhanced decohesion (HEDE) and adsorption-induced decohesion (AIDE), the main fracture lines of ferritic steels are parallel to \{100\} and \{110\}-planes with \{100\}-planes preferred \cite{Du2024, Barik2023, Varanasi2022}. Note, that \{110\} planes have the highest interlayer distance, causing cleavage due to a reduced cohesive strength. 
The directionality along \{100\} planes, on the other hand, can be explained by the much lower diffusion barriers through (100) and type-\textbf{I} surfaces, in general. These \say{easy} entry points into the metal accelerate embrittlement and crack propagation along type-\textbf{I} planes while other orientations effectively inhibit hydrogen ingress.

\section*{Acknowledgements}
This work has been performed as part of the NOHENTRY project, funded by the Energy Transition Fund (Energietransitiefonds) of the Directorate-General Energy (Algemene Directie Energie) of the Federal Public Service for the Economy (FOD Economie) of Belgium. Discussions with Robin Dedoncker, Krista Van Den Bergh and Lode Duprez of OCAS NV are gratefully acknowledged. L.M. acknowledges financial support from the Research Foundation - Flanders (FWO) through a fundamental research fellowship [grant no. 1199325N] S.C. acknowledges financial support from OCAS NV by an OCAS-endowed chair at Ghent University. The computational resources and services used in this work were provided by the VSC (Flemish Supercomputer Center), funded by the FWO and the Flemish Government - department EWI.

\section*{Conflict of Interest}
The authors declare no conflict of interest.

\newpage

\printbibliography

\appendix
\section{Supplementary Information}

\subsection{Vicinal Surface Geometry}\label{appx:geometry}

One way to visualize the geometry of different surface orientation (or crystallographic planes) is by stereographic projection. This mathematical tool is used to correlate different surface orientations in a crystal lattice, uniquely defined by their angular relationship to the crystal axes, to a point on a two-dimensional map.
To create a stereographic projection, one takes the points on the hull of a sphere where the surface normals intersect with the hull of a sphere that is placed around the bcc crystal in its origin. Per convention, the poles of the sphere are aligned with the primary crystal orientations, i.e., with the [100] normal in case of a cubic crystal. As each surface normal intersects the sphere twice and because different equivalent surfaces correspond to different surface normals (i.e., (100), (010), and (001)), only a small part of the whole spherical surface is sufficient to represent all possible surface orientations of a cubic crystal. In this reduced representation, the stereographic projection has the shape of a triangle with (per convention) the low-index surfaces (100), (110), and (111) as the edge points. All other surface orientations fall naturally within the area enclosed by these points on the spherical surface.
Geometrically speaking, the circular arc connecting the (100) and (110) edge gives a measurement of the azimuth angle enclosed by the surface normal and the [100] direction up to an angle of 45$\degree$, which defines the (110) orientation. Taking the (110) direction as a starting point and continuously increasing the polar angle up to 45$\degree$, one follows the circular arc up to the edge defining the (111) orientation. 

The exact position of a point of a certain surface orientation within the stereographic projection can is determined by its miller indices $(h, k, l)$. These determine the direction of the surface normal and describe the angular relationship of its surface normal with the crystal orientations.  
Assuming (100) as the reference direction, the azimuth angle $\varphi$ (and thus, the projection's position on the horizontal arc) can be determined by a conversion into polar coordinates as
\begin{eqnarray}
    \tan(\varphi) = \frac{k}{h}.
\end{eqnarray}

Equally, the polar angle $\vartheta$, which corresponds to the angle enclosed between the surface normal and the [001]-direction, is determined by 

\begin{eqnarray}
    \cos(\vartheta) = \frac{l}{\sqrt{h^2+k^2+l^2}}.
\end{eqnarray}

\newpage
\subsection{Computational Details}\label{appx:comp_details}

\begin{table}[ht]
    \centering
    \begin{tabular}{cccccc}
    \toprule
    Surface (hkl) & x[$a_0$] & y[$a_0$] & z[\AA] &\# frozen layers& \#Fe atoms \\ 
    \midrule
    (100)     &$2$          & $2$         & 11.33 & 3 & 36  \\
    (110)     &$\sqrt{2}$  & 2           & 12.02 & 3 & 28  \\
    (111)     &$\sqrt{6}$   & $\sqrt{2}$  & 11.45 & 4 & 30  \\
    (210)     & $\sqrt{5}$  & $2$         & 12.03 & 6 & 40  \\
    (211)     &$2\sqrt{2}$  & $\sqrt{3}$  & 13.87 & 3 & 52  \\
    (221)     &$3\sqrt{2}$  & $\sqrt{2}$  & 12.74 & 7 & 56  \\
    (410)     &$\sqrt{17}$  & 2           & 12.71 & 9 & 76  \\
    (411)     &$3$          & $\sqrt{2}$  & 11.35 & 5 & 36  \\
    (441)     &$\sqrt{66}$  & $\sqrt{2}$  & 11.34 & 12 & 94  \\
    (421)     &$\sqrt{105}$ & $\sqrt{5}$  & 12.67 & 10 & 210 \\ 
    \bottomrule
    \end{tabular}
    \caption{Comparison of the unit cell sizes of the modeled surface orientations. The lateral unitcell dimensions (x, y) of the examined surfaces are indicated in units of $a_0$ = 2.832 \AA~and the vertical dimensions (z) are given in \AA. The last columns indicates the number of frozen layers of iron atoms to model the bulk and total number of iron atoms required to model a specific surface unitcell.}
    \label{tab:unitcell_measures}
\end{table}

\begin{table}[ht]
    \centering
    \begin{tabular}{lcccc}
    \toprule
        Surface (hkl)& $x^{PES}$ [$a_0$] & $y^{PES}$ [$a_0$] & \#x (d [Å]) & \#y (d [Å]) \\ \hline
        \midrule
        (100) & $1/2$          & $1/2$        & 6 (0.28)  & 6 (0.28)\\ 
        (110) & $1/\sqrt{2}$   & 1/2          & 5 (0.5)   & 3 (0.71) \\ 
        (111) & $\sqrt{6}$     & $1/\sqrt{2}$ & 13 (0.58) & 5 (0.5)\\
        (211) & 1/$\sqrt{2}$   & $\sqrt{3}/2$ & 5 (0.5)   & 7 (0.41) \\
        (411) & $3$             & $1/\sqrt{2}$ & 19 (0.47) & 5 (0.5) \\ 
        \hline
        (221) & 9/2$^{a}$    & $1/\sqrt{2}$ & 19 (0.71) & 5 (0.5) \\ 
        (410) & 4$^{b}$     & 1  & 17 (0.71) & 5 (0.71)\\ 
        (441) & 2.5$^{c}$       & $1/\sqrt{2}$ & 11 (0.74) & 5 (0.5) \\ 
        (210) & 3/$\sqrt{2}^{d}$ & 1/2   & 11 (0.6) & 3 (0.71) \\
        (421) & 3/$\sqrt{2}^{d}$ & 5$^{e}$ & 7 (1.0)& 11 (1.42) \\ 
    \bottomrule
    \end{tabular}
    \caption{Summary of the sampling parameters for the PESs. The first two columns indicate the sampling area of the PES. Therein, we provide the lateral measurements of the PES sampling area in units of $a_0=2.832$~Å. For the (100), (110), (111), (211), and (411) surface orientations, the sampling takes place parallel to the respective surface orientation. For the remaining vicinal surfaces the sampling takes place parallel to the respective low index planes. In the last two columns, the number of sampling points considered along the respective direction and the lateral distance between two points (in parentheses) are given. The superscripts in the first column indicate PES sampling parallel to the indicated vicinal planes.
    \footnotesize{$a:$ along the [$\bar{1}10$] direction, $b:$ along [$00\bar{1}$], $c:$ along [$110$], $d:$ along [$1\bar{1}0$], $e:$ along [$100$]}}
    \label{tab:PES_sampling}
\end{table}
\end{document}